%% file: HQ-TLP.tex
\documentclass[conference]{IEEEtran}
\IEEEoverridecommandlockouts
\usepackage{cite}
\usepackage{amsmath,amssymb,amsfonts}
\usepackage{algorithmic}
\usepackage{graphicx}
\usepackage{textcomp}
\usepackage{xcolor}
\def\BibTeX{{\rm B\kern-.05em{\sc i\kern-.025em b}\kern-.08em
    T\kern-.1667em\lower.7ex\hbox{E}\kern-.125emX}}
\begin{document}

\title{Towards a Unified Method for Network Dynamic via Adversarial Weighted Link Prediction
}

\author{\IEEEauthorblockN{Meng Qin}
\IEEEauthorblockA{\textit{Independent Research}\\
mengqin\_az@foxmail.com}
}

\maketitle

\begin{abstract}
Network dynamic (e.g., traffic burst in data center networks and channel fading in cellular WiFi networks) has a great impact on the performance of communication networks (e.g., throughput, capacity, delay, and jitter). This article proposes a unified prediction-based method to handle the dynamic of various network systems. From the view of graph deep learning, I generally formulate the dynamic prediction of networks as a temporal link prediction task and analyze the possible challenges of the prediction of weighted networks, where link weights have the wide-value-range and sparsity issues. Inspired by the high-resolution video frame prediction with generative adversarial network (GAN), I try to adopt adversarial learning to generate high-quality predicted snapshots for network dynamic, which is expected to support the precise and fine-grained network control. A novel high-quality temporal link prediction (HQ-TLP) model with GAN is then developed to illustrate the potential of my basic idea. Extensive experiments for various application scenarios further demonstrate the powerful capability of HQ-TLP.
\end{abstract}


\section{Introduction}
High dynamic (e.g., in terms of traffic and topology) is a significant feature of modern communication networks and considered as one of the major bottlenecks of system performance. One possible reason is that conventional network control methodologies are based on current system states. In these methodologies, only when current states (e.g., traffic and topology) are exactly observed or measured, the control decisions can be eventually made based on the observed states. However, the high dynamic of system states makes the time interval from (\romannumeral1) \textit{the epoch that current system states can be completely observed} to (\romannumeral2) \textit{the epoch that control decision is required to be made} quite limited. Some sophisticated but time-consuming techniques (e.g., artificial intelligence (AI) models) cannot be applied to existing network systems due to the stringent real-time requirements.

The prediction of system states (e.g., mobility, traffic, and topology) based on systems' historical behaviors has emerged as a promising solution to break the performance bottleneck. For instance, in data center networks (DCN), traffic prediction techniques could be used to schedule the parallel network flows, so as to avoid the performance degradation due to resource shortages \cite{Nie2017Traffic}. For mobile edge computing (MEC), users' location can be forecasted to guide the resource allocation on edge servers for the sake of better Quality of Services (QoS) \cite{Plachy2016Dynamic}. 
%In the Internet of Vehicle (IoV), the prediction of the distance between mobile node and base station can effectively support the intelligent handoff of heterogeneous environments (e.g., LTE and WiMAX).
In summary, the primary motivation of this article is that \textit{if the network dynamic can be accurately predicted, the control decision can be made based on the prediction result to pre-allocate key resources before the exact future network states are completely observed}. It gives the system additional time to apply more sophisticated network control techniques, reaching the potential to meet with more complicated application requirements.

Although numerous prediction-based techniques have been developed to tackle network dynamic \cite{Nie2017Traffic,Plachy2016Dynamic}, their manually selected features and designed models are limited to specific application scenarios (e.g., traffic prediction in DCN), which are hard to be generalized to other scenarios.

In this article, I focus on a unified method to handle the network dynamic of various scenarios from the view of graph deep learning. For various complex systems (e.g., social and communication networks), graphs serve as a generic model to describe entities and their relations using sets of nodes and edges. For example, in DCN, one can model each host (i.e., entity) as a node. The traffic between a pair of hosts (i.e., relation) can be modeled as an edge between this node pair. Namely, given the defined entities and their relations, the system behavior in a certain time step can be represented as a graph snapshot. The network dynamic can be described by a sequence of such snapshots in multiple successive time steps. Under this setting, the prediction-based solution to network dynamic can be generally formulated as the temporal link prediction (TLP) task \cite{qin2023temporal}. Given the graph snapshots in previous time steps, TLP aims to predict next snapshot, which indicates the future system behavior to support the prediction-based network control.

In the past several decades, researchers have discussed the TLP applications in social networks (e.g., friend recommendation in social media) and biological networks (e.g., evolution analysis of cancer). However, few of them considered the application in communication networks. Furthermore, most existing TLP approaches mainly focus on the prediction of unweighted snapshots (i.e., determine the existence and absence of edges) but substantially ignore the informative edge weights. In fact, edge weights can be used to quantitatively describe some system states (e.g., delay, flow, signal strength, and distance) beyond unweighted relations. The prediction of weighted snapshots should not only determine the existence of edges but also estimate corresponding edge weights. It remains a challenging task that most existing TLP techniques cannot tackle \cite{Lei2019GCN,qin2023high,qin2023temporal}.

This article serves as a continuation of my prior work \cite{Lei2018Adaptive,Lei2019GCN,qin2023high}. To the best of my knowledge, this article is the first:
\begin{itemize}
    \item (\romannumeral1) to discuss a unified weighted TLP method to handle the dynamic of various communication network systems;
    \item (\romannumeral2) to generate high-quality weighted snapshots with adversarial learning (e.g., GAN) to support the precise and fine-grained network control.
\end{itemize}
In the rest of this article, I first conclude a unified method to model system behaviors and analyze the challenges of weighted TLP in Section~\ref{Solution}. In particular, I discuss the potential to generate high-quality predicted snapshots with adversarial learning. In Section~\ref{Model}, a novel high-quality temporal link prediction (HQ-TLP) model is developed as an example to illustrate the superiorities of my basic idea. The effectiveness of HQ-TLP is further evaluated by conducting extensive experiments on the datasets of various scenarios in Section~\ref{Exp}. Finally, Section~\ref{Con} concludes this article and indicates future prospects.

\section{The Unified Prediction-Based Method}\label{Solution}
\subsection{Modeling Network Dynamic}
\begin{figure}[htbp]
\centerline
{\includegraphics[width=\linewidth, trim = 15 15 15 15, clip]{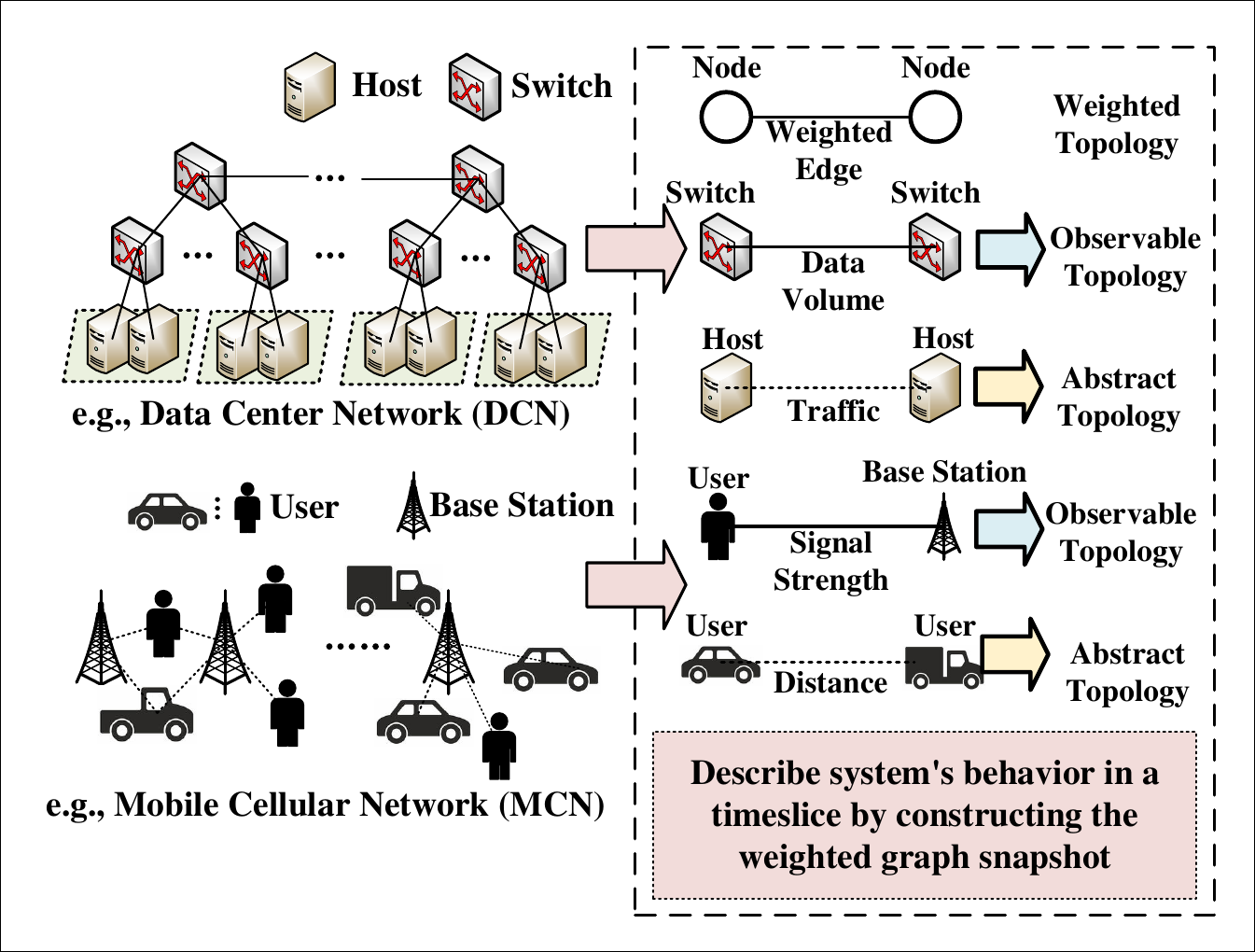}}
\caption{Illustration of the unified strategy to model system behaviors in a time step by constructing a weighted graph snapshot.}
\label{Conc}
\end{figure}

In general, the behaviors of various communication network systems in a specific timeslice can be abstracted as a corresponding weighted graph snapshot. As demonstrated in Fig.~\ref{Conc}, the abstraction of a snapshot is two-fold. 

On the one hand, \textit{the snapshot topology can be directly extracted from the observable or measurable physical links of a system}. For instance, in DCN, one can model each switch as the node and extract the topology from the physical links between switches (e.g., the fat-tree topology). The data volume transmitted on the link between a pair of switches can be described as the link weight. In mobile cellular networks (MCN), one can model user equipments (UEs) and base stations (BSs) as heterogeneous nodes. If a UE establishes a measurable link to a BS, there is an edge between this node pair, where the link signal strength can be defined as the corresponding edge weight.

On the other hand, \textit{the snapshot topology can also be indirectly constructed based on some abstract quantities of the system}. For example, in DCN, one can treat each host server as a node and consider the traffic between each pair of hosts as the corresponding weighted edge, where traffic is an abstract high-level quantity transmitted via the observable switch topology.
Furthermore, in MCN, a snapshot can be constructed based on the distance between UE nodes. If the distance between a pair of UEs is smaller than a certain threshold, there is an edge between this node pair, whose weight could be inversely proportional to the distance (i.e., smaller distance indicates larger weight).

The topology of a graph snapshot with $n$ nodes can be formally described by an adjacency matrix ${\bf{A}}$, which is an $n$-dimensional symmetry square matrix. The element in the $i$-th row and $j$-th column of ${\bf{A}}$ describes the edge weight between the $i$-th and $j$-th nodes in this snapshot. Especially, the zero elements in ${\bf{A}}$ indicates that there is no edge between corresponding node pairs. The network dynamic can be further described by a sequence of snapshots that describes the system behavior in multiple successive time steps.

As a simple demonstration of the proposed unified method to handle network dynamic, I consider the case that all the snapshots share a common node set. I further discuss some more complicated applications and their possible solutions in the last section of this article.

\subsection{High-Quality Weighted TLP}
Given the sequence of historical snapshots, TLP aims to derive the possible snapshot in the next time step. Such a prediction result indicates the possible future system behavior, which can be used to pre-allocate key system resources (e.g., caches, channels, and CPU time). However, there remain the following challenges \cite{qin2023high,qin2023temporal} for the prediction of weighted topology that existing techniques may fail to handle. These challenges make the prediction-based method hard to support the precise and fine-grained network control.

\textbf{The Wide-Value-Range Issue} \cite{qin2023high,qin2023temporal}. In a certain graph snapshot, edge weights may have a large value range (e.g., from $0$ to $20,000$) depending on the concrete systems and modeling strategies. In particular, there may be a non-ignorable proportion of edges that have relatively small weights. For most TLP techniques, the basic training criterion is to \textit{minimize the reconstruction error} between the prediction result and ground-truth. This criterion is only sensitive to edges with large weights, failing to reflect the magnitude difference between small weights. For example, the magnitude difference between $1$ and $2$ is significantly larger than that between $1,990$ and $2,000$, even though the latter may result in larger error. For the prediction-based network control, failing to distinguish the magnitude difference between small edge weights implies that \textit{the system tends to pre-allocate much more resources for some links than their actual demands}, which results in unnecessary waste of resources.

\textbf{The Sparsity Issue} \cite{qin2023high,qin2023temporal}. Most of the extracted graph snapshots are also sparse, in which there are non-ignorable pairs of system entities without the defined relations. For most TLP methods, prediction results are given in the form of adjacency matrices. In an adjacency matrix, a zero element means there is no edge between the corresponding node pair. A non-zero element with small value indicates there is still an edge between this node pair but the weight is small. These two cases are entirely different. However, distinguishing between the two cases with entirely different physical meanings remains  challenging for existing methods based on the \textit{minimization of reconstruction error}.
Failing to distinguish between small edge weights and zero values implies that the prediction may mistakenly direct the system to (\romannumeral1) \textit{allocate resources for the nonexistent links} or (\romannumeral2) \textit{not allocate resources for existing links}, which may also lead to the waste of system overhead.

\begin{figure}[htbp]
\centerline
{\includegraphics[width=\linewidth, trim = 45 30 30 25, clip]{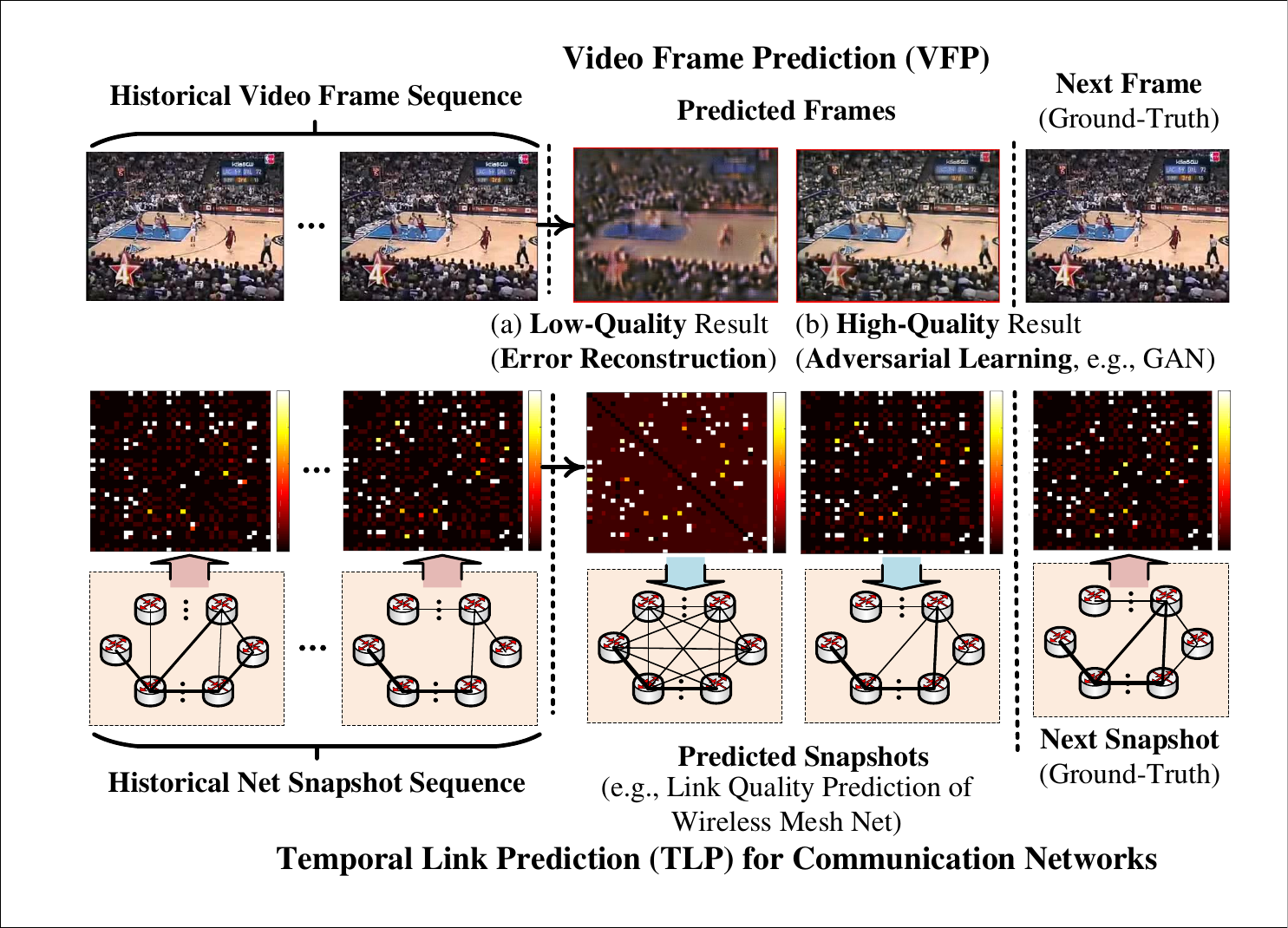}}
\caption{Illustration of VFP and TLP, as well as the comparison between (a) low-quality prediction via error reconstruction and (b) high-quality prediction via adversarial learning, with the figure adapted from \cite{qin2023high,qin2023temporal}.}
\label{TLP}
\end{figure}

The utilization of GAN \cite{Goodfellow2014Generative} can be considered as a possible solution to the aforementioned challenges. GAN is an emerging AI technique with several promising applications in computer vision (CV). The example of video frame prediction (VFP) \cite{Mathieu2016Deep} is shown in the top-left side of Fig.~\ref{TLP}. Similar to TLP, VFP aims to predict the next frame image based on historical frames. Minimizing the reconstruction error is also the mainstream training strategy of related approaches but the corresponding predicted frame is usually with low quality in terms of resolution (see (a) in Fig.~\ref{TLP}). In contrast, the prediction result generated via GAN has significantly higher resolution and is much closer to the ground-truth (see (b) in Fig.~\ref{TLP}).
A typical GAN framework consists of a generator $G$ and a discriminator $D$, which compete in a minimax game. On the one hand, \textit{$D$ tries to distinguish the real data in the training set from the data generated by $G$}. On the other hand, \textit{$G$ tries to fool $D$ by generating plausible data}. For VFP, the adversarial process can significantly enhance the ability of $G$ to generate high-resolution frame images.

Inspired by high-resolution VFP, I consider the potential to generate high-quality graph snapshots by introducing the adversarial process. In the bottom-left side of Fig.~\ref{TLP}, the adjacency matrices of all the snapshots w.r.t. the prediction of one example in \textit{UCSB}, a dataset about the link quality of a mesh network (see Table~\ref{Data}), are visualized. To highlight the difference between zeros and small edge weights, I set all the zeros in each adjacency matrix to be $-200$. In this way, black represents the zero value (set to be $-200$). The color close to dark red means small edge weights while that close to white indicates large values. I develop an example adversarial prediction model named HQ-TLP (see Section~\ref{Model} for its details) and compare its prediction result with that given by the reconstruction-based LSTM model. In Fig.~\ref{TLP}, the reconstruction-based result (i.e., case (a)) can only fit large edge weighs but fails to distinguish differences between zeros and small weights. On the other hand, the result of HQ-TLP (i.e., case (b)) is with high quality, fitting small, large, and zero weights well. The comparison preliminarily demonstrates the potential of adversarial learning to handle the \textit{sparsity} and \textit{wide-value-range} issues.

\section{An Example Adversarial Prediction Model}\label{Model}
\begin{figure}[htbp]
\centerline
{\includegraphics[width=\linewidth, trim = 25 20 25 15, clip]{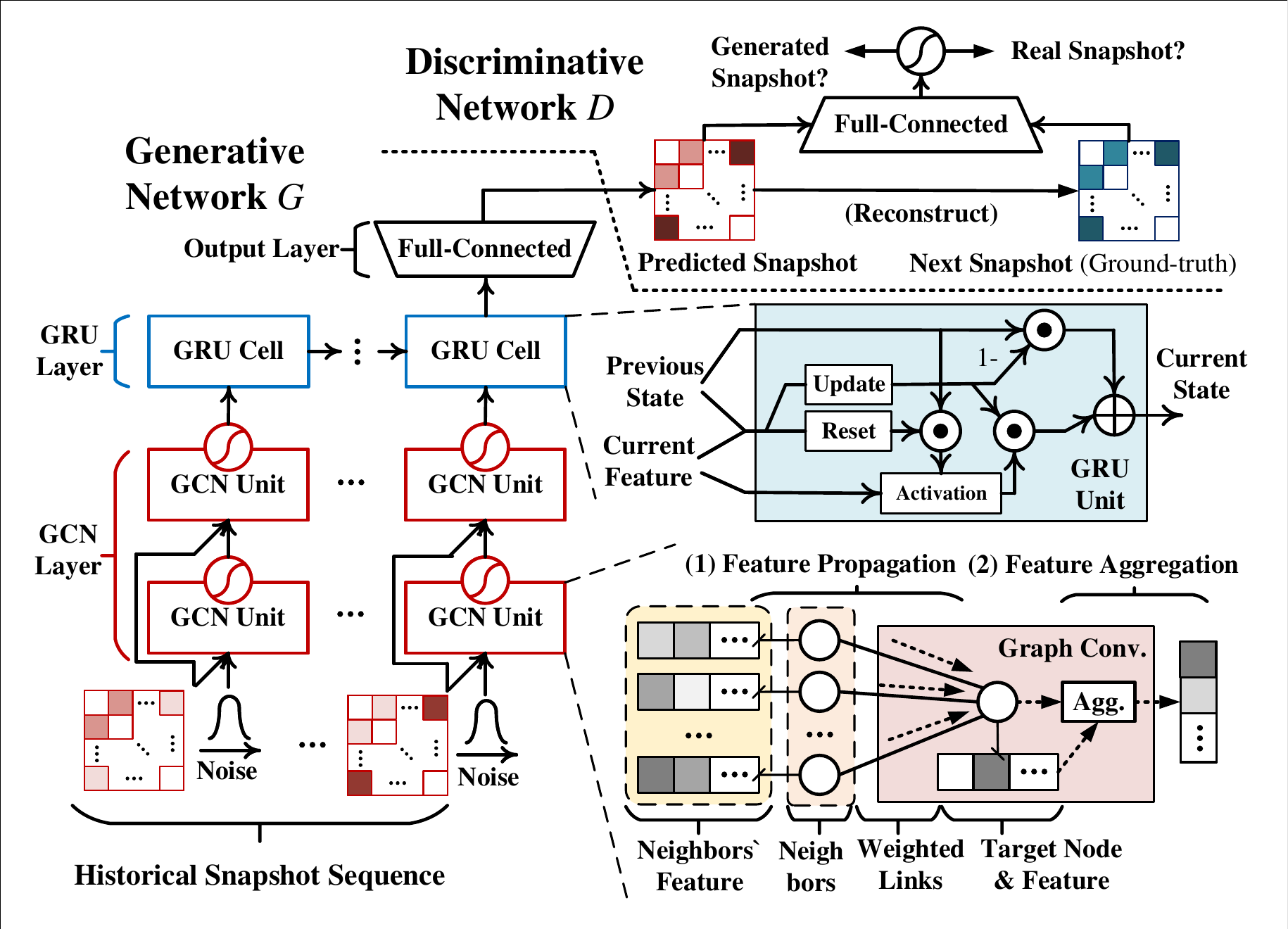}}
\caption{The architecture of HQ-TLP, with a generative network $G$ and a discriminative network $D$.}
\label{Arch}
\end{figure}

As an example of the aforementioned discussion, I develop HQ-TLP by leveraging (\romannumeral1) GAN \cite{Goodfellow2014Generative}, (\romannumeral2) graph convolutional network (GCN) \cite{Kipf2017Semi-supervised} and (\romannumeral3) gated recurrent unit (GRU) \cite{Chung2014Empirical}. Fig.~\ref{Arch} presents the major architecture of HQ-TLP.

It is based on the GAN framework, including a generative network $G$ and a discriminative network $D$. In HQ-TLP, $D$ is a fully-connected feedforward neural network (a.k.a. multi-layer perceptron). $G$ has a more sophisticated architecture with two GCN layers, one GRU layer, and one fully-connected output layer. In the rest of this section, I in sequence introduce the components of GAN, GCN, and GRU. For simplicity, I omit their mathematics details that can be found in \cite{Goodfellow2014Generative,Kipf2017Semi-supervised,Chung2014Empirical}.

\subsection{GAN}\label{GAN}
When the system comes to a new time step, the generative network $G$ is used to predict the next snapshot. It takes (\romannumeral1) historical adjacency matrices and (\romannumeral2) a series of random noises as inputs. The discriminative network $D$ is trained to enhance the capability of $G$ to give high-quality prediction results via an adversarial process. For $D$, the prediction ground-truth (from the training dataset) and generated snapshot (from $G$) are alternatively treated as the input.

In the original GAN framework \cite{Goodfellow2014Generative}, $G$ is trained based on the unique objective to generate samples that can successfully fool $D$. However, it may also output plausible samples entirely different from the ground-truth. For instance, in CV, GAN is verified to be capable to generate plausible new images that are not in the training set \cite{Radford2016Unsupervised}. The generation of such plausible data is inappropriate for TLP, since each predicted snapshot is expected to be consistent with the ground-truth. To address this limitation, I introduce an additional constraint regarding the ground-truth. Concretely, the objective of $G$ is two-fold:
\begin{itemize}
    \item (\romannumeral1) minimizing the error between the generated snapshot and ground-truth; 
    \item (\romannumeral2) fooling $D$ with the refined generated result.
\end{itemize}
The first objective is the \textit{error reconstruction criterion} of conventional TLP techniques. It directs $G$ to generate a snapshot consistent with ground-truth. The second objective is the optimization strategy of GAN, which enhances the ability of $G$ to generate high-quality prediction results. I formulate these two criteria into one objective by linearly combining their objective functions.
Similar to the standard GAN, the objective of $D$ is:
\begin{itemize}
    \item to distinguish the ground-truth from the generated result given by $G$.
\end{itemize}
The objectives of $D$ and $G$ jointly form a minimax adversarial game, enabling the model to tackle the \textit{wide-value-range} and \textit{sparsity} issues. For a given initialized state, the joint optimization of $G$ and $D$ can be implemented via gradient descent, where one can alternatively take the following two steps until convergence:
\begin{itemize}
    \item (\romannumeral1) updating the parameters of $D$  based on its optimization objective with $G$ fixed;
    \item (\romannumeral2) updating the parameters of $G$ based on its optimization objective with $D$ fixed.
\end{itemize}

\subsection{GCN}\label{GCN}
In $G$, the two GCN layers are used to capture the topological features of each single snapshot. Different from convolutional neural network (CNN) applied to images or text, GCN is the variant for graphs. It is originally designed for the semi-supervised node classification problem, where each node has attribute features in addition to topology structures. Hence, network topology and node attributes are two input sources of GCN.
As illustrated in Fig.~\ref{Arch}, a typical GCN unit takes node features as input and conducts the \textit{graph convolution operation} based on graph topology. The final output is generated in the same way with a standard fully-connected layer, where each node is assigned with another aggregated feature. The \textit{graph convolution process} on an arbitrary target node in the graph includes the following two steps.
\begin{itemize}
    \item (\romannumeral1) Neighbors of a target node \textit{propagate} their features to the target node.
    \item (\romannumeral2) The target node \textit{aggregates} the propagated features and its own feature via a certain function (e.g., the weighted average based on edge weights) and derives a new aggregated feature.
\end{itemize}
In this way, the key characteristics of graph topology can be captured in the aggregated output. Moreover, the GCN output can also be used as the attribute input of a new GCN unit, forming a multi-layer structure.

In HQ-TLP, I only consider the network topology but treat the noise input of GAN as a special form of node attributes. For each time step, the corresponding adjacency matrix is the common input of the two GCN layers. As shown in Fig.~\ref{Arch}, the first GCN layer takes the random noise as (attribute) input and then fed the aggregated attributes to the second layer. In particular, the multi-layer GCN also has the potential to explore and capture some deep structural properties (e.g., high-order proximities and community structures \cite{qin2018adaptive,qin2019towards,li2019identifying,qin2021dual,gao2023raftgp}) beyond the observed graph topology.

\subsection{GRU}\label{GRU}
GRU is an advanced recurrent neural network (RNN), which is effective to learn the long-term dependencies of sequential data. In HQ-TLP, GRU is applied to explore the evolving patterns of multiple successive snapshots. As depicted in Fig.~\ref{Arch}, GRU can be described as an encapsulated cell with several gate units, including a \textit{reset gate} (RG), an \textit{update gate} (UG), and a \textit{candidate activation gate} (CAG). For each time step, the GRU cell takes the (\romannumeral1) current topological features (from GCN) and (\romannumeral2) GRU state in the previous time step as inputs. It then outputs the GRU state of current time step. Particularly, the current feature input describes the new content of network dynamic while the previous GRU state serves as a comprehensive description of the past content.

The three gate units in GRU play different roles. RG allows GRU to forget previous states according to the comprehensive effect of two input sources. CAG includes new features of current time step into GRU. UG eventually decides how much the GRU updates its state to mix the new and previous content. Therefore, the historical features and current content of network can be comprehensively preserved.

Finally, I treat the last output of GRU w.r.t. the sequence input of $G$ as the representation of all the historical snapshots and feed it into the output layer to derive a prediction result.

\subsection{The Prediction Algorithm}
\begin{figure}[htbp]
\centerline
{\includegraphics[width=0.99\linewidth, trim = 20 20 20 25, clip]{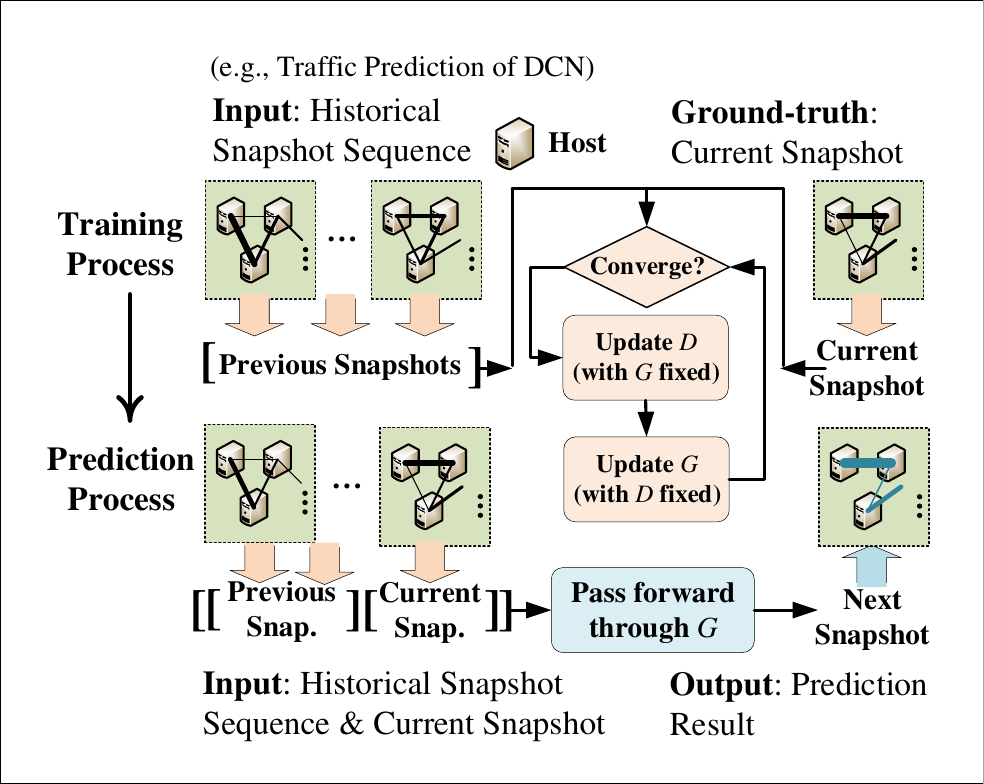}}
\caption{Algorithm to train HQ-TLP and generate the prediction result with traffic prediction of DCN as an example.}
\label{Alg}
\end{figure}

For the optimization of HQ-TLP, I adopt the reasonable assumption that \textit{the snapshot close to the next time step should have more similar characteristics to the ground-truth (i.e., the next snapshot) compared with those far from it}. It is also a criterion of commonly adopted by most existing TLP techniques. When the system comes to a new time step, I first conduct the training process to update model parameters with historical snapshots as the input and current snapshot as the ground-truth. In this way, the trained model can capture the latest evolving patterns of network because current snapshot (i.e., training ground-truth) is considered to be most similar to the next snapshot (i.e., prediction ground-truth). After training the model, I conduct the prediction process to generate the prediction result with historical snapshot sequence (including current snapshot) as input. I summarize the aforementioned procedures in Fig.~\ref{Alg} with the traffic prediction in DCN as an example application.

\section{Extensive Experiments}\label{Exp}
\subsection{Datasets of Various Scenarios}
I apply HQ-TLP to three datasets covering different communication network scenarios. Descriptions of the datasets are shown in Table~\ref{Data}, where $N$ and $T$ are the numbers of nodes and time steps, respectively.

\begin{table}[htbp]
\caption{Datasets descriptions, with $N$ and $T$ as the numbers of nodes and time steps.}
\label{Data}
\begin{tabular}{l|lll|l}
\hline
Datasets & $N$ & $T$ & Wei. Range & Scenarios \\ \hline
\textit{UCSB} & 38 & 1,000 & $[0, 2000]$ & Wireless Mesh Networks \\
\textit{KAIST} & 92 & 500 & $[0, 250]$ & Mobile Ad-Hoc Networks \\
\textit{NumFab} & 128 & 350 & $[0, 20000]$ & Data Center Networks \\ \hline
\end{tabular}
\end{table}

\textit{UCSB} \cite{Ramachandran2007Routing} is a router link quality dataset collected from a real wireless mesh network, with measurable links between wireless routers. Hence, I adopted the first strategy in Section~\ref{Solution} to extract the graph snapshots.  \textit{KAIST} \cite{Rhee2011On} is a real trajectory dataset of user mobility, which forms a mobile ad-hoc network. \textit{NumFab} is a simulated host server flow dataset of DCN generated via the simulation code of \cite{Nagaraj2016NUMFabric}. According to the second strategy in Section~\ref{Solution}, I constructed snapshots based on the (\romannumeral1) distance between each UE pair for \textit{KAIST} and (\romannumeral2) traffic volume between each host pair for \textit{NumFabric}.

\subsection{Metrics for Weighted TLP}
I quantitatively evaluated HQ-TLP by using the metrics of (\romannumeral1) root mean square error (RMSE), (\romannumeral2) edge-wise KL-divergence (EW-KL), and (\romannumeral3) mismatch rate (MR). In particular, I proposed EW-KL and MR in my prior work \cite{Lei2019GCN}.

RMSE is a commonly used metric for most TLP techniques. It measures the error between the adjacency matrices given by the prediction result and ground-truth. Note that RMSE is based on the conventional \textit{reconstruction error criterion}. As discussed in Section~\ref{Solution}, it cannot reflect the \textit{wide-value-range} and \textit{sparsity} properties of weighted networks.

To further measure whether a TLP method can give high-quality prediction results, I used EW-KL and MR. EW-KL evaluates the magnitude difference between edge weights of the prediction result and ground-truth by computing the KL-divergence between corresponding elements in adjacency matrices. Note that there may exist the zero exception for KL-divergence, where one of the corresponding edge weights in the prediction result or ground-truth is zero. Such exception corresponds to the \textit{sparsity} issue in Section~\ref{Solution}. I ignored the zero exception for EW-KL and specially calculated the proportion of such `mismatched' edges in a snapshot, which is defined as MR.

In general, smaller RMSE, EW-KL, and MR indicate better prediction quality. For more mathematical details, please refer to my prior work \cite{Lei2019GCN}.

\subsection{Baseline Methods}
I compared HQ-TLP with six TLP baselines, including CN-SVD \cite{Ma2017Nonnegative}, CN-NMF \cite{Ma2017Nonnegative}, GrNMF \cite{Ma2018Graph}, AM-NMF, \cite{Lei2018Adaptive}, LSTM, and GRU.

CN-SVD and CN-NMF are matrix factorization (MF) methods based on the collapsed network (CN) model. In the CN model, successive snapshots are first linearly combined as a comprehensive static snapshot, which is defined as the CN. Some MF techniques (e.g., singular value decomposition (SVD) and non-negative matrix factorization (NMF)) can then be applied to CN. GrNMF and AM-NMF are NMF-based methods without collapsing the dynamic network. They directly generate prediction results by solving their unified models that integrate historical graph topology. Note that these four approaches are not data-driven. When the sequence of historical snapshots is given, they can directly generate prediction results by taking the following two steps:
\begin{itemize}
    \item (\romannumeral1) learning a low-dimensional representation by combing historical snapshots while considering that \textit{the snapshot close to the next time step should be more similar to the ground-truth compared with those far from it};
    \item (\romannumeral2) constructing the prediction result by conducting the inverse process of MF based on the learned representation (i.e., recovering from the learned hidden space to the original topology space).
\end{itemize}

In contrast, LSTM and GRU represent data-driven methods that directly utilize LSTM and GRU with the \textit{error reconstruction criterion}. Different from the first four MF-based methods, the data-driven approaches (including HQ-TLP) may take effect (i.e., automatically learning the latent knowledge regarding network dynamic) only when the model is fully trained on a certain amount of data.

\subsection{Performance Evaluation}
\begin{figure}[htbp]
\centerline
{\includegraphics[width=\linewidth, trim = 22 25 22 20, clip]{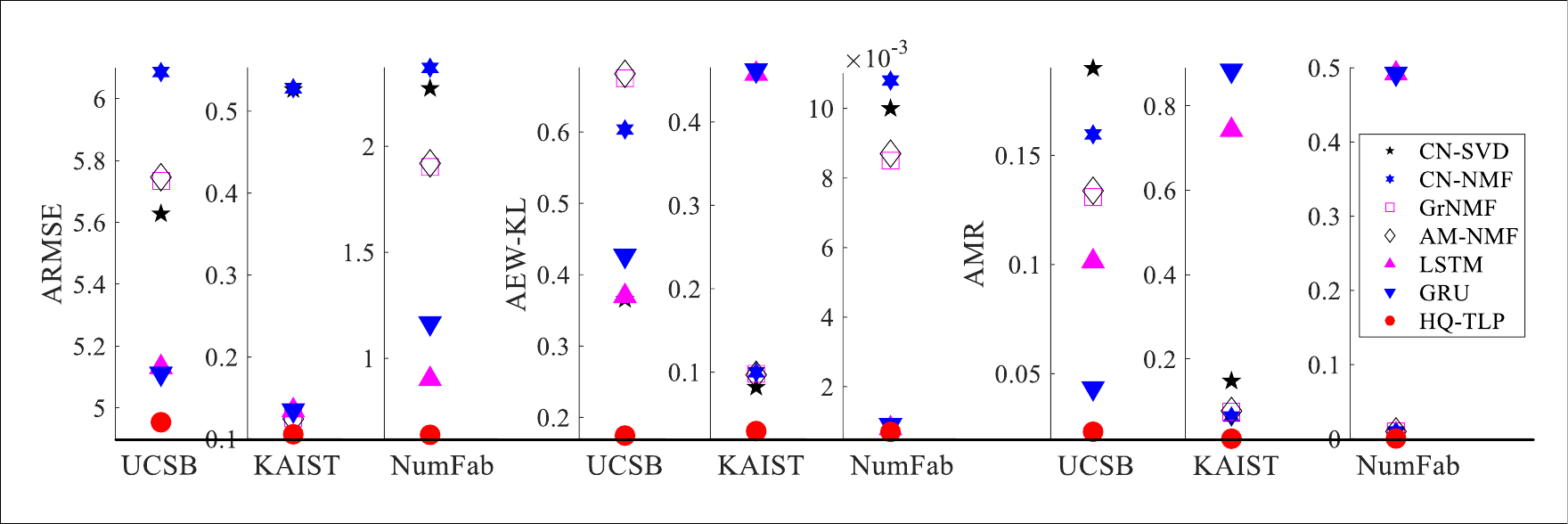}}
\caption{Evaluation experiment results with ARMSE, AEW-KL, and AMR of the last 50 snapshots on test sets.}
\label{Perf-Eva}
\end{figure}

In experiments, I set the length of the sequence of historical snapshots to be $10$. For all the methods, I measured the performance of the last $50$ snapshots (i.e., the test set) on each dataset, with the rest snapshots (i.e., the training set) employed to train data-driven models. Results of average RMSE (ARMSE), average EW-KL (AEW-KL), and average MR (AMR) on test sets are depicted in Fig.~\ref{Perf-Eva}.

According to Fig~\ref{Perf-Eva}, HQ-TLP has the best ARMSE, AEW-KL, and AMR on all the datasets. An example of the high-quality prediction snapshot on \textit{UCSB} (in terms of adjacency matrix) is also shown in Fig.~\ref{TLP} (see case (b)), where HQ-TLP can fit large, small, and zero weights well. The experiment results comprehensively indicate that the adversarial TLP is effective to handle the \textit{wide-value-range} and \textit{sparsity} issues of weighted dynamic networks. It can be a possible core technique to support the precise and fine-grained network control in future communication network systems. Especially, the prediction-based control has the significant potential to break the performance bottleneck of conventional current-state-based control, dealing with the high dynamic of systems.

\section{Conclusion and Future Prospects}\label{Con}
In this article, I proposed a unified prediction-based network control methodology and formulated the dynamic prediction of various scenarios as the challenging TLP task for weighted dynamic networks. An adversarial model HQ-TLP is developed as an example to handle the \textit{wide-value-range} and \textit{sparsity} issues, generating high-quality prediction results. Extensive experiments demonstrated the potential of HQ-TLP to tackle the high dynamic of modern communication networks while supporting the precise and fine-grained prediction-based network control.
Although my approach demonstrates great potential, there remain the following concerns that need further discussions in future work.

\textbf{Dynamic of Node Sets}. In this article, I assumed that all the network snapshots share a common node set. However, system dynamic also indicates that there may exist (\romannumeral1) \textit{new entities added into} and (\romannumeral2) \textit{old entities deleted from the system}. A possible solution is to combine TLP and active node prediction (ANP) into a multi-task learning scheme. Every system has its service ceiling, which can be used to estimate the maximum possible number of nodes in a network. ANP aims to determine whether a potential node has links to other nodes in a snapshot (i.e., in an active or inactive state), given the past node states. By sharing the temporal information with TLP, ANP can be used to predict the set of active nodes in the next snapshot, where the deleted entities are represented by inactive nodes. Moreover, the inductive nature of deep graph learning \cite{qin2023high,qin2023towards}, which aims to directly generalize a trained model to new unseen topology, can also be considered to tackle the dynamic of node sets.

\textbf{Prediction Error}. The prediction-based control has the potential to handle network dynamic but there remains the prediction error. Although conventional current-state-based control techniques need to meet with the real-time constraints, it can observe exact network states. A promising research direction is to leverage these two methods, where the prediction-based approach can be applied to refine the coarse-grained control decision given by a conventional method. Furthermore, to determine the lower bound of prediction error w.r.t. a given scenario is also a significant theoretical problem. It can be used to quantitatively measure the potential and risk of the prediction-based control (i.e., to what extend the prediction-based method can tackle the network dynamic).

\textbf{Snapshot Sampling}. In experiments, I uniformly sampled all the snapshots. In fact, the sampling interval of snapshots may also affect prediction results. More experiments should be conducted to further explore such an effect. Since the system dynamic may change over time, an adaptive sampling strategy is a better alternative to network control. For instance, the sampling frequency should be relatively high in rush hours to accurately reflect the high dynamic of a system. In contrast, it can be relatively low when the system is not very busy to minimize system overheads.

\ifCLASSOPTIONcaptionsoff
  \newpage
\fi

\input{HQ-TLP.bbl}

\bibliographystyle{IEEEtran}
%\bibliography{HQ-TLP.bib}

\end{document}